\documentclass[a4paper,12pt]{article}
\usepackage[utf8]{inputenc}
\usepackage{tabularray}
\usepackage{amsmath,amsthm,amsbsy,amsfonts,comment,booktabs}
\usepackage{bm,graphicx,psfrag,epsf,booktabs,bbm,multirow}
\usepackage{enumitem}
\usepackage{amsmath}
\usepackage[toc,page]{appendix}
\usepackage[round, authoryear]{natbib}
\usepackage{url} 
\usepackage{caption}
\usepackage{subcaption}
\usepackage{ifpdf}
\usepackage{listings}
\usepackage{setspace}
\usepackage{stmaryrd}
\RequirePackage[OT1]{fontenc}
\usepackage{algorithm}
\usepackage{algorithmic}

\usepackage[margin=1in]{geometry}

\usepackage{amsbsy}

\usepackage[mathscr]{eucal}

\usepackage{bm}

\usepackage{xspace}

\mathchardef\mhyphen="2D

\ifpdf
\usepackage[colorlinks,urlcolor=blue,citecolor=blue,linkcolor=blue]{hyperref}
\else
\newcommand{\href}[2]{{#2}}
\usepackage{url}
\fi

\ifpdf
\newcommand{\Sec}[1]{\hyperref[sec:#1]{Section~\ref*{sec:#1}}} 
\newcommand{\App}[1]{\hyperref[sec:#1]{Appendix~\ref*{sec:#1}}} 
\newcommand{\Supp}[1]{\hyperref[sec:#1]{Supplement~\ref*{sec:#1}}} 
\newcommand{\Eqn}[1]{\hyperref[eq:#1]{{\rm (\ref*{eq:#1})}}} 
\newcommand{\Part}[1]{\hyperref[part:#1]{(\ref*{part:#1})}} 
\newcommand{\Fig}[1]{\hyperref[fig:#1]{Figure~\ref*{fig:#1}}} 
\newcommand{\Tab}[1]{\hyperref[tab:#1]{Table~\ref*{tab:#1}}} 
\newcommand{\Thm}[1]{\hyperref[thm:#1]{Theorem~\ref*{thm:#1}}} 
\newcommand{\Lem}[1]{\hyperref[lem:#1]{Lemma~\ref*{lem:#1}}} 
\newcommand{\Prop}[1]{\hyperref[prop:#1]{Proposition~\ref*{prop:#1}}} 
\newcommand{\Obs}[1]{\hyperref[ob:#1]{Observation~\ref*{ob:#1}}} 
\newcommand{\Cor}[1]{\hyperref[cor:#1]{Corollary~\ref*{cor:#1}}} 
\newcommand{\Def}[1]{\hyperref[def:#1]{Definition~\ref*{def:#1}}} 
\newcommand{\Alg}[1]{\hyperref[alg:#1]{Algorithm~\ref*{alg:#1}}} 
\newcommand{\Ex}[1]{\hyperref[ex:#1]{Example~\ref*{ex:#1}}} 
\newcommand{\As}[1]{\hyperref[as:#1]{Assumption~{\rm\ref*{as:#1}}}} 
\newcommand{\Reg}[1]{\hyperref[as:#1]{Condition~\ref*{reg:#1}}} 
\newcommand{\AlgLine}[2]{\hyperref[alg:#1]{line~\ref*{line:#2} of Algorithm~\ref*{alg:#1}}}
\newcommand{\AlgLines}[3]{\hyperref[alg:#1]{lines~\ref*{line:#2}--\ref*{line:#3} of Algorithm~\ref*{alg:#1}}}
\else
\newcommand{\Sec}[1]{{Section~\ref{sec:#1}}} 
\newcommand{\App}[1]{{Appendix~\ref{sec:#1}}} 
\newcommand{\Supp}[1]{{Supplement~\ref{sec:#1}}} 
\newcommand{\Eqn}[1]{{(\ref{eq:#1})}} 
\newcommand{\Part}[1]{{(\ref{part:#1})}} 
\newcommand{\Fig}[1]{{Figure~\ref{fig:#1}}} 
\newcommand{\Tab}[1]{{Table~\ref{tab:#1}}} 
\newcommand{\Thm}[1]{{Theorem~\ref{thm:#1}}} 
\newcommand{\Lem}[1]{{Lemma~\ref{lem:#1}}} 
\newcommand{\Prop}[1]{{Proposition~\ref{prop:#1}}} 
\newcommand{\Cor}[1]{{Corollary~\ref{cor:#1}}} 
\newcommand{\Def}[1]{{Definition~\ref{def:#1}}} 
\newcommand{\Alg}[1]{{Algorithm~\ref{alg:#1}}} 
\newcommand{\Ex}[1]{{Example~\ref{ex:#1}}} 
\newcommand{\Reg}[1]{{R~\ref*{reg:#1}}} 
\fi

\newcommand{\Real}{\mathbb{R}}

\newtheorem{observation}{Observation}[section]
\newtheorem{theorem}{Theorem}[section]

\newtheorem{definition}{Definition}[section]
\newtheorem{corollary}{Corollary}[section]

\newenvironment{tightcenter}{%
  \setlength\topsep{0pt}
  \setlength\parskip{0pt}
  \begin{center}
}{%
  \end{center}
}
\begin{document}

\def\spacingset#1{\renewcommand{\baselinestretch}%
{#1}\small\normalsize} \spacingset{1.55}
\title{\bf \Large A Stability Framework for Parameter Selection in the Minimum Covariance Determinant Problem }
\author{\textbf{Qiang Heng\footnote{School of Mathematics, Southeast University}\; \footnote{Departments of Computational Medicine, UCLA},\hspace{2mm} Hui Shen\footnote{Department of Mathematics and Statistics, McGill University},\hspace{2mm} Kenneth  Lange\footnote{Departments of Computational Medicine, Human Genetics, and Statistics, UCLA}}}
\date{}
\maketitle
\begin{abstract}
The Minimum Covariance Determinant (MCD) method is a widely adopted tool for robust estimation and outlier detection. In this paper, we introduce MCD model selection based on the notion of stability. Our best subset method leverages prior best practices such as statistical depths for initialization and concentration steps for subset refinement. Our contribution lies in constructing a bootstrap procedure to estimate the instability of the best subset algorithm. The instability path offers insights into a dataset's inlier/outlier structure and facilitates suitable choice of the subset size.  We rigorously benchmark the proposed framework against existing MCD variants and illustrate its practical utility on several real-world datasets.
\end{abstract}

%
\begin{tightcenter}
\end{tightcenter}


\maketitle


%

\section{Introduction}
\label{s:intro}

The minimum covariance determinant (MCD) estimator \citep{rousseeuw1985multivariate} and its subsequent extensions have been widely adopted for robust estimation and outlier detection. This estimator identifies a subset of a predetermined size from an $n \times p$ multivariate data matrix, aiming for the smallest possible determinant of the reduced sample covariance matrix. MCD gained popularity following the introduction of the computationally efficient fast minimum covariance determinant (FastMCD) algorithm \citep{rousseeuw1999fast}. FastMCD has found broad applications in fields such as finance, econometrics, engineering, the physical sciences, and biomedical research \citep{hubert2008high,hubert2018minimum}. The MCD estimator is statistically consistent and asymptotically normal \citep{butler1993asymptotics,cator2012central}.  Recent extensions to the MCD paradigm include a kernelized version \citep{schreurs2021outlier} for non-elliptical data and a cell-wise version \citep{raymaekers2023cellwise} for robustness against cell-wise outliers.

In the univariate case \citep{rousseeuw2005robust}, the MCD problem can be exactly solved in $O(n\log n )$ arithmetic operations. Unfortunately, computation becomes more challenging in the multivariate setting. FastMCD is a greedy block minimization algorithm that provides a locally optimal solution for the non-convex combinatorial optimization problem of minimizing the subsample covariance determinant. Due to its greedy nature, FastMCD can be trapped in local minima. Therefore, proper initialization is crucial, particularly when the proportion of outliers or the dimension of the data increases. Naive random initialization falters in these circumstances. To address this, \cite{hubert2012deterministic} proposed a deterministic initialization strategy called deterministic minimum covariance determinant (DetMCD) that relies on six different initializations for $\mu$ and $\Sigma$. This ensemble strategy outperforms random initialization in robustness when the proportion of outliers is high.

Recent work by \cite{zhang2023fast} suggests that using the trimmed subset induced by the notion of projection depth \citep{zuo2000general} is conceptually simpler, computationally faster, and even more robust. The robustness is likely due to the fact that, asymptotically, projection depth and MCD induce the same elliptical decision region \citep{zuo2000structural}. It is noteworthy that projection depth is equivalent to the Stahel-Donoho outlyingness index \citep{Stahell981breakdown,donoho1982breakdown}, normalized to the range of $[0,1]$ via a monotonic decreasing transform. \cite{hubert2005robpca} first leveraged Stahel-Donoho outlyingness for the purpose of robust principal component analysis. 

To its detriment, MCD relies on the computation of Mahalanobis distances, which requires the invertibility of the sample covariance matrix $\hat{\Sigma}$. To tackle the high-dimensional settings where $p>n$, \cite{boudt2020minimum} introduced Tikhonov regularization to ensure the positive definiteness of $\hat{\Sigma}$. Other tactics for high-dimensional outlier detection explore alternative definitions of Mahalanobis distance and formulate the underlying problem as hypothesis testing. For example, \cite{ro2015outlier}, \cite{li2022outlier}, and \cite{li2024outlier} studied the null asymptotic distribution of test statistics under appropriate model assumption. \cite{hubert2005robpca} employed robust principal components (ROBPCA) to categorize outliers based on their relationship to the principal subspace.  This method distinguishes two types of outliers: high score-distance (SD) points, which are detectable within the principal  subspace, and high orthogonal-distance (OD) points, whose projections onto the principal subspace may be indistinguishable from those of inliers.

A crucial hyperparameter in the MCD problem is the subset size $h$, or equivalently, the inlier proportion $\alpha=\frac{h}{n}$. Selecting an appropriate $h$ has long been a challenging task. Often $h$ is chosen conservatively to ensure the exclusion of every conceivable outlier from the selected subset. Common choices include $h=\lfloor 0.5n \rfloor$ and $h=\lfloor 0.75n \rfloor$. These arbitrary choices potentially compromise statistical efficiency. Various reweighting procedures \citep{rousseeuw1999fast,hardin2005distribution,riani2009finding,ro2015outlier,li2024outlier}  can be applied to rescue additional observations as inliers \citep{rousseeuw1999fast}, but they require the specification of a significance level that balances Type-I error and power. Selecting the appropriate significance level or cut-off threshold complicates model selection. Furthermore, reweighting relies on parametric approximations of the Mahalanobis distances. Such approximations can be unreliable when $p$ approaches $n$ or when the covariance matrix is nearly singular and highly ill-conditioned.

This paper introduces a framework for selecting the appropriate subset size, $h$, or estimating the number of outliers based on the concept of stability. Our best subset selection method integrates established best practices, such as using projection depth for initialization and concentration steps for refining the subset. Inspired by the idea of clustering instability \citep{wang2010consistent,fang2012selection}, we propose a bootstrap procedure to estimate the instability of our $h$-subset selection algorithm. Statistical stability has proven to be effective in selecting regularization parameters \citep{sun2013consistent,wen2023stability} and controlling false discoveries \citep{meinshausen2010stability,wang2011random}. To the best of our knowledge, this is the first application of stability in outlier detection. To select a suitable $h$, the general principle is to identify the subset size that exhibits minimal instability through a grid search. In high-dimensional scenarios where $p > n$, we extend our framework by incorporating robust PCA for dimensionality reduction. Our real-world data applications demonstrate the value of instability in identifying the number of principal components, $q$, that best distinguishes between low-SD and high-SD points.

\section{Background}\label{sec:background}
Given a multivariate data matrix $X\in \Real^{n\times p}$, assume that most  observations are sampled from a unimodal distribution with mean $\mu\in \Real^p$ and covariance $\Sigma\in \Real^{p\times p}$. Additionally, suppose there exists a subset of outliers that significantly deviates from the primary mode. For the purpose of outlier detection and robust estimation, we are interested in finding a subset $H\subset \{1,2,\dots,n\}$ of size $h$ that is outlier-free. The mean and covariance estimated from the subset are given by 
\begin{align}
\begin{split}\label{eq:meancov}
\hat{\mu} & =  \frac{1}{h}\sum_{i\in H} x_i \quad \text{and} \quad
\hat{\Sigma}  =   \frac{1}{h} \sum_{i\in H} (x_i - \hat{\mu})(x_i - \hat{\mu})^\top,
\end{split}
\end{align}
where $x_i\in \Real^p$ denotes the $i$-th row of $X$. 
\begin{definition}\label{def:mcd}
The minimum covariance determinant problem seeks the $h$ observations from $x_1,x_2,\dots,x_n$ minimizing the determinant of $\hat{\Sigma}$ defined by equation \Eqn{meancov}.
\end{definition}
\cite{rousseeuw1999fast} proposed the first computationally efficient algorithm (FastMCD) to tackle the MCD problem. Starting from a random initial subset $H$, FastMCD first estimates $\mu$ and $\Sigma$ via equation \Eqn{meancov}. Then, based on the Mahalanobis distances
\begin{eqnarray}\label{eq:mahalanobis}
d_i & = & \sqrt{(x_i - \hat{\mu})^\top \hat{\Sigma}^{-1}(x_i - \hat{\mu})},
\end{eqnarray}
the observations $x_i, i = 1,2,\dots, n$ are ranked from closest to furthest. In the sequel, we write $\text{MD}(x_i;\hat{\mu},\hat{\Sigma})$, or simply $\text{MD}(x_i)$, to denote the Mahalanobis distance \Eqn{mahalanobis}. Given the permutation $\pi$ producing the ranking $d_{\pi(1)}\le d_{\pi(2)}\le \dots\le d_{\pi(n)}$, $H$ is updated as
\begin{eqnarray}\label{eq:updateH}
H & = & \{\pi(1),\pi(2),\dots,\pi(h)\}.
\end{eqnarray}
\cite{rousseeuw1999fast} referred to the combination of procedures \Eqn{meancov}, \Eqn{mahalanobis}, and \Eqn{updateH} as a concentration step (C-step). Based on a uniqueness property \citep{grubel1988minimal}, \cite{rousseeuw1999fast} showed that each C-step monotonically decreases $\det(\hat{\Sigma})$. 
\subsection{Statistical Depth}\label{sec:depth}

Statistical depth is a nonparametric index commonly used to rank multivariate data from  the center outward \citep{zuo2000general,zhang2023fast}. A statistical depth function increases with the centrality of an observation, taking values between 0 and 1. After computing the statistical depth of all observations within a dataset, it is natural in estimating means and covariances to retain the $h$ observations with the greatest depths.  \cite{zhang2023fast} investigated the application of two representative depth variants, projection depth and $L_2$ depth \citep{zuo2000general}. Their experiments demonstrate that projection depth is more robust across different simulation settings. Additionally, projection depth offers the advantage of affine invariance \citep{zuo2000general,trimming}.  For these reasons, we focus on projection depth to facilitate initialization of MCD.  We again stress that projection depth is simply the Stahel-Donoho outlyingness index \citep{Stahell981breakdown,donoho1982breakdown} normalized to the range of [0,1].

\begin{definition}
The projection depth of a vector $x\in \Real^p$ with respect to a distribution $F$ is defined as 
\begin{equation}\label{eq:projdepth}
D(x;F) = \left [ 1+ \underset{\lVert u\rVert=1}{\sup} \frac{|u^\top x-{\rm med}(u^\top y)|}{{\rm MAD}(u^\top y)}\right ]^{-1},
\end{equation}
where $y$ is a random vector that follows the distribution $F$, ${\rm med}(V)$ denotes the median of a univariate random variable $V$, and ${\rm MAD}(V) = {\rm med}(|V-{\rm med}(V)|)$ is the median absolute deviation from the median. 
\end{definition}
Since a closed-form expression for the quantity \Eqn{projdepth} does not exist, projection depth is typically approximated by generating $k$ random directions $u$ and taking the supremum. For the purpose of ranking the observations from a center outward, one can compute $D(x_i;\hat{F}_n)$ for $i$ between $1$ and $n$, where $\hat{F}_n$ is the empirical distribution of $X\in\Real^{n\times p}$. In this case projection depth is also referred to as sample projection depth. We write $D(x_i;\hat{F}_n)$ as $D(x_i;X)$ to highlight its dependence on the observed data matrix $X$. Sample projection depths are efficiently computed with a time complexity of $O(nkp)$. There are generally two simple methods for generating the random directions appearing in equation (\ref{eq:projdepth}). The naive approach is to sample uniformly from the unit sphere. The second, a more data-driven method, generates random directions as the differences between randomly selected pairs of observations, followed by normalization to unit length \citep{hubert2005robpca}. This tactic is more effective when outliers are concentrated in specific directions or when the dimension is high. To combine the strengths of both approaches, we adopt a hybrid strategy in this paper: the first 500 directions are generated using the data-driven method, while the remaining directions are sampled uniformly from the unit sphere. Our software implementation for computing projection depths is an adaptation of the C++ code in the R package \texttt{ddalpha} \citep{lange2014fast,pokotylo2016depth}. Unless otherwise specified, we generate a total $\max\{1000,100p\}$ random directions. 

\subsection{Reweighted Estimators}\label{sec:reweight}
Many MCD algorithms employ reweighting to avoid excluding too many observations. For example, the fast depth-based (FDB) algorithm of \cite{zhang2023fast} proceeds as follows: (a) computing statistical depths and defining the initial $h$-subset to be the $h$ observations with the largest depths (b) computing $\hat{\mu}$ and $\hat{\Sigma}$ by equation \Eqn{meancov}, and (c) re-estimating $\hat{\mu}$ and $\hat{\Sigma}$ via the reweighting scheme \Eqn{reweighting} of \cite{rousseeuw1999fast}. In summary,
\begin{align}
\begin{split}\label{eq:reweighting}
c  & = \underset{i}{\text{med}} \quad \frac{\text{MD}^2(x_i;\hat{\mu},\hat{\Sigma})}{\chi^2_{p,0.5}}, \quad
w_i   = \begin{cases} 1 & \text{MD}^2(x_i;\hat{\mu},c\hat{\Sigma}) \le \chi^2_{p,0.975}\\
0 & \text{MD}^2(x_i;\hat{\mu},c\hat{\Sigma}) > \chi^2_{p,0.975}, 
\end{cases} \\
\hat{\mu}_{\text{re}} & = \frac{\sum_{i=1}^n w_i x_i}{\sum_{i=1}^n w_i}, \quad
\hat{\Sigma}_{\text{re}}  =  \frac{\sum_{i=1}^n w_i (x_i - \hat{\mu}_{\text{re}})(x_i - \hat{\mu}_{\text{re}})^\top}{\sum_{i=1}^n w_i -1}.
\end{split}
\end{align} FDB amends the depth-induced $h$-subset to include observations with weight $1$ and skips concentration steps entirely. However, in practice, a reweighting procedure like \Eqn{reweighting} with a subjectively chosen cut-off value can lead to significant underestimation or overestimation of the number of outliers. 

\subsection{ROBPCA}
In the under-determined case where $p>n$, the naive MCD estimator is stymied by the singularity of the estimated covariance matrix. One popular alternative in high-dimensional outlier detection is ROBPCA \citep{hubert2005robpca}. Specifically, given robust estimates of the center $\hat{\mu}$ and scatter $\hat{\Sigma}$, along with the robust principal components $P\in \Real^{p\times q}$ (the eigenvectors of $\hat{\Sigma}$), the robust principal component scores are computed as:
\begin{eqnarray*}
Z & = & (X-1_n \hat{\mu}^\top)P,
\end{eqnarray*}
where $1_n$ is an $n \times 1$ column vector of ones.

\cite{hubert2005robpca} then defines two types of distances to measure the outlyingness of each observation. The first is the score distance (SD)
\begin{equation}
\text{SD}_i = \sqrt{\sum_{j-1}^q \frac{z_{ij}^2}{l_j}},
\end{equation}
where the $l_j$ are the eigenvalues of $\hat{\Sigma}$. The second is the orthogonal distance, given by:
\begin{equation}\label{eq:od}
\text{OD}_i = \lVert x_i - \hat{\mu} - P z_i \rVert_2,
\end{equation}
where $z_i\in \Real^q$ is the $i$-th row of $Z$. Both high-SD and high-OD points are considered outliers. To detect high SD points, one can leverage the fact that the squared score distances follow a $\chi^2_q$ distribution. For the squared orthogonal distances, the distribution must be approximated using parameters estimated from the data. Specifically, it has been shown that the orthogonal distances to the power of 2/3 are approximately normally distributed. After computing the orthogonal distances, we may apply univariate MCD to the numbers $OD_i^{2/3}$ with a conservative subset size (for example, $h=\lfloor 0.5n\rfloor$) to estimate a robust mean $\hat{\mu}_{OD}$ and a robust variance $\hat{\sigma}^2_{OD}$. Points having an OD index greater than the cutoff $(\hat{\mu}_{OD}+\hat{\sigma}_{OD}z_{0.975})^{3/2}$ are declared outliers. While one can select low-SD points nonparametrically by applying the MCD algorithm to the principal component scores $Z$, to the best of our knowledge, no existing methods nonparametrically select high-OD points. 

\section{Methods}\label{sec:methods}
\subsection{Best Subset Selection}

Our algorithm for selecting an inlier subset of a given subset size $h$ combines prior best practices, namely projection depth initialization \citep{hubert2005robpca,zhang2023fast} and concentration steps \citep{rousseeuw1999fast}.  In contrast, the FDB algorithm of \cite{zhang2023fast}, which skips concentration steps, is able to work well despite possible outliers in the initial $h$-subset because the reweighting step \Eqn{reweighting} adds an additional layer of projection. One of the main objectives of this paper is to present a nonparametric alternative to reweighting, so we choose to omit reweighting in our subset selection algorithm. 
\begin{algorithm}[htbp]
\caption{Depth-initialized Minimimum Covariance Determinant (Depth-MCD)}\label{alg:depthsubset}
\begin{algorithmic}[1]
\REQUIRE Data matrix $X\in \Real^{n\times p}$, subset size $h<n$, number of random directions $k$ (default is $\max\{1000,100p\}$).
\STATE Compute the the projection depths $d_i=D(x_i;X),i=1,2,\dots,n$, using $k$ random directions.
\STATE Sort $d_i$ in decreasing order, yielding permutation $\pi$ with $d_{\pi(1)}\ge d_{\pi(2)}\ge\dots \ge d_{\pi(n)}$. Set $H=\{\pi(1),\pi(2),\dots,\pi(h)\}$.
\STATE Update $H$ using concentration steps \Eqn{meancov}, \Eqn{mahalanobis}, and \Eqn{updateH} until $H$ stabilizes.
\ENSURE The best $h$-subset $H$
\end{algorithmic}
\end{algorithm}

\subsection{Model Selection via Stability}\label{sec:noutlier}
In the MCD literature, selecting the right subset size $h$ poses a consistent challenge. Drawing inspiration from the notion of \textit{clustering instability} \citep{wang2010consistent,fang2012selection}, we now describe a bootstrap procedure for estimating the number of inliers $h$ (equivalently the number of outliers $n-h$). Outlier detection can be framed as clustering into two groups: inliers and outliers. Unlike traditional clustering, the outlier cluster does not necessarily exhibit spatial structure. This spatial ambiguity causes classic strategies in clustering, such as the gap statistic \citep{tibshirani2001estimating}, to fail.  The rationale for using stability as the instrument is that if the number of inliers $h$ is suitably chosen, then different bootstrapped samples should yield similar conclusions about which cases are outliers in the data matrix $X$. Suppose the bootstrapped samples $\dot{X}$ and $\ddot{X}$ deliver indicator functions (binary maps) $\psi_1$ and $\psi_2$ distinguishing outliers from inliers, with a value of $1$ denoting an outlier and a value of $0$ denoting an inlier. To measure the distance between $\psi_1$ and $\psi_2$, one can resort to 
the clustering distance:
\begin{definition}\label{def:clusterdis}
The clustering distance between two clusterings $\psi_1$ and $\psi_2$ on the data matrix $X$ is  
\begin{equation}
d_{X}(\psi_1,\psi_2) = \frac{1}{n^2} \sum_{i=1}^n \sum_{j=1}^n |I_{\psi_1(x_i)=\psi_1(x_j)}-I_{\psi_2(x_i)=\psi_2(x_j)}|,
\end{equation}
where $\psi_1$ and $\psi_2$ each map a vector $x\in \Real^p$ to a cluster label $k\in \{0,1,\dots,K-1\}$.
\end{definition}
In general, computing $d_{X}(\psi_1,\psi_2)$ as indicated in \Def{clusterdis} takes $O(n^2)$ operations. For the two-cluster case ($K=2$), this simplifies to the following computationally efficient formula. 
\begin{theorem}\label{thm:distance}
When there are only two clusters $(K=2)$, the clustering distance is given by
$$
d_X(\psi_1,\psi_2) = 2p_{X}(\psi_1,\psi_2)[1-p_{X}(\psi_1,\psi_2)],
$$
where $p_{X}(\psi_1,\psi_2) = \frac{1}{n} \sum_{i=1}^n |\psi_1(x_i)-\psi_2(x_i)|$.
\end{theorem}
Proof of \Thm{distance} is provided in the supplement. In principle, both $d_X$ and $p_X$ are viable candidates for use as instability metrics when performing inlier/outlier classification. In this paper, we favor $p_X$ due to its simplicity. 

\cite{haslbeck2020estimating} observed that clustering distance can be adversely affected by differences in cluster sizes. In our context, $p_X(\psi_1,\psi_2)$ becomes problematic when $h$ approaches $n$, because at this point, there is little room for the binary mappings $\psi_1$ and $\psi_2$ to differ. For example, in the extreme situation where $h=n$, $p_X(\psi_1,\psi_2)$ will always be $0$. To adjust for the effect of cluster sizes in $p_X(\psi_1,\psi_2)$, we consider the scaled metric: 
\begin{equation}\label{eq:estimatedcorrected}
p_{X}^c(\psi_1,\psi_2) = \frac{1}{c}p_{X}(\psi_1,\psi_2)
\end{equation}
where $c = 2\frac{h}{n}\frac{(n-h)}{n}$ is the expectation of $|\psi_1(x_i)-\psi_2(x_i)|$ when $\psi_1$ and $\psi_2$ are two independent random binary clusterings that each assign a random set of $h$ out of the $n$ data points to one cluster and the remaining $n - h$ to the other. To measure the overall instability of the algorithm,  we apply a  $\log(x+1)$ transformation on \Eqn{estimatedcorrected} and average  over $B$ independent bootstrap pairs
\begin{equation}\label{eq:estimates}
\hat{s}(h) = \frac{1}{B} \sum_{i=1}^B \log(1+p_{X}^c(\psi_{1,b},\psi_{2,b})).
\end{equation}
Here, the binary maps $\psi_{1,b}$ and $\psi_{2,b}$ are subscripted by the bootstrap pair $b$.  The $\log(x+1)$ transformation ensures consistency with the same transformation applied later to the Wasserstein distance. 

\begin{algorithm}[htbp]
\caption{Instability Estimation}\label{alg:insta}
\begin{algorithmic}[1]
\REQUIRE Data matrix $X\in \Real^{n\times p}$, subset size $h<n$, number of bootstrap pairs $B$. 
\FOR{$b=1:B$}
    \STATE Construct a pair of bootstrapped samples $\dot{X}_b$ and $\ddot{X}_b$.
    \STATE Apply \Alg{depthsubset} to $\dot{X}_b$ and $\ddot{X}_b$ to obtain best $h$-subsets $\dot{H}_b$ and $\ddot{H}_b$. 
    \STATE Compute Mahalanobis distances $\dot{d}_i = \text{MD}(x_i;\dot{\mu},\dot{\Sigma})$ and $\ddot{d}_i = \text{MD}(x_i;\ddot{\mu},\ddot{\Sigma})$ for $i\in [n]$, where $(\dot{\mu},\dot{\Sigma})$, $(\ddot{\mu},\ddot{\Sigma})$ are estimated from $\dot{H}_b$, $\ddot{H}_b$, respectively.
    \STATE Set $\psi_{1,b}(x_i)=0$ if $\dot{d}_i(x_i)$ is within the $h$-largest and $\psi_{1,b}(x_i)=1$ otherwise. $\psi_{2,b}(x_i)$ is defined similarly.
    \STATE Compute $p_{X}^c(\psi_{1,b},\psi_{2,b})$ by equation \Eqn{estimatedcorrected}.
\ENDFOR
\ENSURE $\hat{s}(h)$ in \Eqn{estimates}.
\end{algorithmic}
\end{algorithm}

\Alg{insta} outlines the workflow of our instability estimation algorithm, using \Eqn{estimatedcorrected} as the dissimilarity measure and \Eqn{estimates} as the instability measure. Although \Alg{insta} may seem computationally intensive at first glance, certain intermediary quantities can be reused during the grid search over $h$. Specifically, we set the depths of observations within a bootstrapped sample to be their computed depth in the original sample. This is motivated by the intuition that bootstrapping generally preserves the relative outlyingness of each observation, or at least it provides a sufficiently accurate approximation for the purpose of initializing concentration steps. The same projection depths also apply for all values of $h$. While it is also possible to use projection depths instead of Mahalanobis distances in step 4 of \Alg{insta}, this would significantly increase the computational cost.

\subsection{Combatting Masking Outliers}
Although the true subset size typically exhibits a high degree of stability, stable separation of inliers and outliers may also occur when large outliers mask intermediate ones. This can lead to the instability path exhibiting multiple minima or stable points, making it unreliable to solely rely on the minimum clustering instability point to determine the true subset size.  To address this challenge, we introduce an additional notion of instability to complement the limitations of clustering instability in dealing with masking outliers.

Specifically, we consider the Wasserstein distance between the normal distributions $\mathcal{N}(\dot{\mu},\dot{\Sigma})$ and $\mathcal{N}(\ddot{\mu},\ddot{\Sigma})$, where $(\dot{\mu},\dot{\Sigma})$ and $(\ddot{\mu},\ddot{\Sigma})$ are robust estimates of $(\mu,\Sigma)$ obtained by applying MCD with subset size $h$ to the bootstrap pairs $\dot{X}_b$ and $\ddot{X}_b$. The squared Wasserstein distance is defined as:
\begin{equation}
\mathcal{W}^2_2(\mathcal{N}(\dot{\mu},\dot{\Sigma}),\mathcal{N}(\ddot{\mu},\ddot{\Sigma})) = \|\dot{\mu} - \ddot{\mu}\|_2^2 + \mathrm{Tr}\left(\dot{\Sigma} + \ddot{\Sigma} - 2 \left(\dot{\Sigma}^{1/2} \ddot{\Sigma} \dot{\Sigma}^{1/2}\right)^{1/2}\right).
\end{equation}
The rationale behind using the Wasserstein distance is that $(\dot{\mu},\dot{\Sigma})$ and $(\ddot{\mu},\ddot{\Sigma})$ should remain similar if the subset is outlier-free but become significantly different once outliers are present. While other metrics involving only $\mu$ or $\Sigma$ (such as $\|\dot{\mu} - \ddot{\mu}\|_2$ or the KL divergence between $\dot{\Sigma}$ and $\ddot{\Sigma}$) could serve a similar purpose, the Wasserstein distance is preferred because it provides a more comprehensive measure of difference that incorporates both $\mu$ and $\Sigma$.

To stabilize the variance of the Wasserstein distances, we apply a $\log(1+x)$ transformation. The resulting metric is defined as:
\begin{eqnarray}
\hat{w}(h) & = & \frac{1}{B}\sum_{b=1}^B \log(1+\mathcal{W}_2(\mathcal{N}(\dot{\mu}_b,\dot{\Sigma}_b),\mathcal{N}(\ddot{\mu}_b,\ddot{\Sigma}_b))).
\end{eqnarray}
While $\hat{w}(h)$ can provide some insight into the true subset size, it tends to underestimates the inlier proportion. Nevertheless, in contrast to clustering instability, $\hat{w}(h)$ is highly sensitive to outliers of large magnitude, making it a valuable complementary metric. 

Combining clustering instability and the Wasserstein distance into a single metric requires careful consideration. Notably, Wasserstein distances are sensitive to both the dimensionality and distributional patterns of the data, which can result in values that are not directly comparable across different scenarios. To address this, it is essential to normalize the Wasserstein distances appropriately. Furthermore, we want clustering instability to remain the dominant component of the metric, with the Wasserstein distances serving as a corrective adjustment. To this end, we propose the following Integrated Instability Metric (IIM):
\begin{eqnarray}\label{eq:final}
\hat{I}(h) & = & (1-\beta) \hat{s}(h) + \beta(\hat{w}(h) - \underset{l}\min \{\hat{w}(l)\}) .
\end{eqnarray}
This final metric is a convex combination of $\hat{s}(h)$ and $\hat{w}(h) - \underset{l}\min \{\hat{w}(l)\}$, with the parameter $\beta \in (0,1)$. The combination parameter $\beta$ is chosen such that the standard deviation of $(1-\beta) \hat{s}(h)$ (over the search grid of $h$) is $\lambda$ times that of $\beta(\hat{w}(h) - \underset{l}\min \{\hat{w}(l)\})$. Based on empirical evidence, we consistently use $\lambda=3$ throughout the paper, as it performs well across a wide range of scenarios. An intuitive justification is that, assuming the two additive components are independent, setting $\lambda=3$ ensures that $(1-\beta) \hat{s}(h)$ accounts for 90\% of the variance in the final metric, while $\beta(\hat{w}(h) - \underset{l}\min \{\hat{w}(l)\})$ contributes the remaining 10\%. This allocation helps prioritize clustering instability while allowing the Wasserstein distance to fine-tune the result.

We note that subtracting $\underset{l}\min \{\hat{w}(l)\}$ from $\hat{w}(h)$ ensures that the log Wasserstein distances have a minimum value of 0. The scale of the log Wasserstein distances is already normalized through multiplication with $\beta$. In practice, if either the location or the scale of the log Wasserstein distances is not properly adjusted, it can lead to undesirable outcomes. Specifically, improper normalization may cause the Wasserstein distances to either dominate the clustering instability in the final metric or introduce a bias towards a smaller number of principal components in \Alg{insta2}, since higher-dimensional spaces naturally lead to larger Wasserstein distances.

\subsection{Extension to ROBPCA}

The previous two sections discussed instability metrics for the MCD estimator in the over-determined $n>p$ case. While ROBPCA provides an effective tool  for classifying outliers relative to the PCA subspace for high-dimensional data, it requires specifying the number of principal components in advance. This number is typically chosen based on the proportion of robust variance explained. In theory, more principal components create more space for high-SD points and leave less room for high-OD points. Thus, the number of principal components creates an important trade-off to be navigated. In this section, we propose a new stability procedure as an alternative. Specifically, we suggest nonparametrically selecting high-SD points by applying the MCD procedure after robust PCA projection. In other words, we apply \Alg{depthsubset} to the principal component scores $Z$ to select the high-SD points. Then the number of principal components $q$ and subset size $h$ are chosen to minimize the instability of subset selection. This tactic enables identification of the parameter settings  that result in the sharpest contrast between low-SD and high-SD points. We demonstrate later that this principle leads to reasonable parameter choices in several real data applications.

\begin{algorithm}[htbp]
\caption{Instability Estimation for ROBPCA}\label{alg:insta2}
\begin{algorithmic}[1]
\REQUIRE Data matrix $X\in \Real^{n\times p}$, subset size $h<n$, number of principal components $q$, number of bootstrap pairs $B$. 
\FOR{$b=1:B$}
    \STATE Construct a pair of bootstrapped samples $\dot{X}_b$ and $\ddot{X}_b$.
    \STATE Perform robust PCA on $\dot{X}_b$ and $\ddot{X}_b$ to obtain robust means and principal components $(\dot{\mu}_b,\dot{V}_b)$ and $(\ddot{\mu}_b,\ddot{V}_b)$, and principal component scores $\dot{Z}_b$ and $\ddot{Z}_b$. 
    \STATE Center and right multiply $X$ by $(\dot{\mu}_b,\dot{V}_b)$ and $(\ddot{\mu}_b,\ddot{V}_b)$ respectively to obtain $\dot{Z}$ and $\ddot{Z}$.  
    \STATE Apply \Alg{depthsubset} to $\dot{Z}_b$ and $\ddot{Z}_b$ to obtain the best subsets $\dot{H}_b$ and $\ddot{H}_b$. 
    \STATE Compute Mahalanobis distances $\dot{d}_i = \text{MD}(z_i;\dot{Z}_b[\dot{H}_b,])$ and $\ddot{d}_i = \text{MD}(z_i;\ddot{Z}_b[\ddot{H}_b,])$ for $i\in[n]$.
    \STATE Set $\psi_{1,b}(x_i)=0$ if $\dot{d}_i$ is within the $h$-largest and $\psi_{1,b}(x_i)=1$ otherwise. Define $\psi_{2,b}(x_i)$ similarly.
    \STATE Compute $p_{X}^c(\psi_{1,b},\psi_{2,b})$ by equation \Eqn{estimatedcorrected}.
\ENDFOR
\ENSURE $\hat{s}(h,q)$ in \Eqn{estimates}.
\end{algorithmic}
\end{algorithm}

To perform robust PCA on the bootstrapped samples,  we leverage the \texttt{PcaHubert} function \citep{hubert2005robpca} from the R package \texttt{rrcov} with $\alpha=0.5$.  The choice of $\alpha=0.5$ prioritizes  robustness and ensures that the center and principal component matrix is free from the influence of outliers. We can similarly compute average log Wasserstein distance $\hat{w}(h,q)$ and the integrated instability metric $\hat{I}(h,q)$, in addition to the clustering instability $\hat{s}(h,q)$.   Once the optimal parameter setting for detecting high-SD points is identified, the parametric test proposed by \cite{hubert2005robpca} can be used to detect high-OD points.

\section{Simulated Examples}\label{sec:ne}

\subsection{Two-dimensional Examples}
We begin with four two-dimensional examples to illustrate the instability paths produced by clustering instability, log Wasserstein distance, and the corresponding integrated instability metric. The outlier-free data consists of $1000\times 2$ standard normal deviates. Next, we consider the following four settings, each with a different pattern of outlier contamination. We apply \Alg{insta} and search over the grid $h=500,525,\dots,975$, using 50 bootstrap pairs for each setting: 
\begin{itemize}
\item Setting 1: 10\% of the observations are replaced with $\mathcal{N}(5,1)$.
\item Setting 2: 10\% of the observations are replaced with $\mathcal{N}(0,15)$ deviates, and another 5\% of the observations are replaced with $\mathcal{N}(0,1000)$ deviates.
\item Setting 3: 10\% of the observations are replaced with $\mathcal{N}(5,1)$ observations, and another 20\% of the observations are replaced with values $(10,10),(20,20),\dots,(2000,2000)$.
\item Setting 4: No outliers are introduced.
\end{itemize}

\begin{figure}[tbp]
  \centering
  \includegraphics[width=\textwidth]{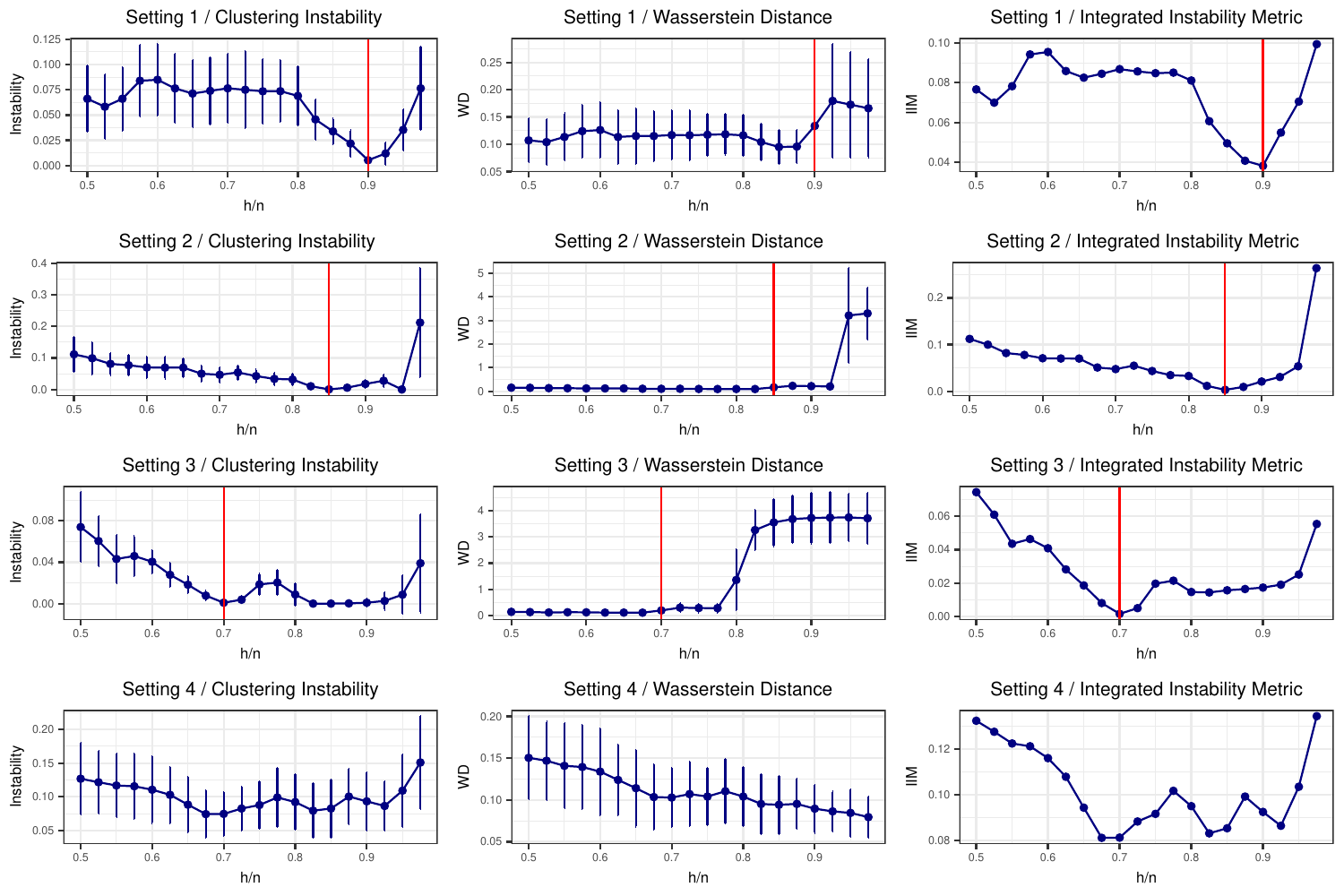}
  \caption{Instability paths for setting 1,2,3, and 4. The left column shows the clustering instability paths, the middle column shows log Wasserstein distances, and the right column shows the integrated instability metrics. For setting 1,2, and 3, the true inlier proportion is highlighted with a red vertical line.}
  \label{fig:2Ds}
\end{figure}

\Fig{2Ds} visualizes the results. We observe that, in general, the true inlier proportion corresponds to a relatively stable point with low clustering instability. However, as discussed earlier and observed in settings 2 and 3, masking outliers can create spuriously stable points with identical or even lower clustering instability. This suggests that selecting the subset size that minimizes clustering instability may result in significant underestimation of the number of outliers.
For the log Wasserstein distance, we observe that it generally plateaus at values lower than the true inlier proportion, but then spikes to a large value once outliers are included. In contrast to clustering instability, masking outliers are easily detected when using log Wasserstein distance as the instability metric. However, log Wasserstein distance typically reaches its minimum before the true subset size. This may be because a bootstrapped sample might contain a larger proportion of outliers than the original sample, leading to the minimization of log Wasserstein distance when the subset size is small enough to exclude all outliers.

We note that, in general, clustering instability nearly resolves the problem but is hindered by the presence of masking outliers. The instability paths produced by the log Wasserstein distance complement this limitation well, and the carefully designed integrated metric \Eqn{final} combines the strengths of both metrics, correctly identifying the true subset size in settings 1, 2 and 3. However, we acknowledge that the integrated metric may slightly underestimate the proportion of inliers due to the incorporation of log Wasserstein distance. Nevertheless, we believe that underestimating the inlier proportion is less problematic than underestimating the outlier proportion. In the supplement, we also demonstrate with simulations that the integrated metric still identifies a subset size that is closer to the true subset size than reweighting using $h=\lfloor 0.5n\rfloor$ as the initial subset size,  especially if $p/n$ is high. 

In the outlier-free setting, extra care is warranted. From the last row of \Fig{2Ds}, we observe that there are generally no significantly stable subset sizes in the clustering instability path. Meanwhile, the log Wasserstein distance exhibits a clear downward trend. When this occurs, log Wasserstein distance path serves as a visual indication that there are no obvious outliers in the data or that the number of inliers exceeds the largest value on the grid.

\subsection{High-dimensional Examples}
In this section, we visualize additional instability paths using the simulation protocol applied in the evaluation of both DetMCD \citep{hubert2012deterministic} and FDB \citep{zhang2023fast}, demonstrating the capability of the instability framework in high-dimensional, complex scenarios. This shared protocol first generates inliers as $x_i=Gy_i$, where $y_i$ is sampled from a Gaussian distribution $\mathcal{N}(0,I_p)$. The $p\times p$ matrix $G$ has diagonal elements equal to 1 and off-diagonal elements equal to 0.75. The outliers are categorized into four types, depending on a scalar $r$ that determines the separation between inlier and outlier clusters.
\begin{itemize}
\item Point outliers: $y_i\sim \mathcal{N}(ra\sqrt{p}, 0.01^2I_p)$, where $\|a\|=1$ and $\sum_i a_i =0$. 
\item Cluster outliers: $y_i\sim \mathcal{N}(rp^{-1/4}1, I_p)$.
\item Random outliers: $y_i\sim \mathcal{N}(rp^{1/4}\nu_i/\lVert \nu_i\rVert, I_p)$, where $\nu_i\sim \mathcal{N}(0,I_p)$.
\item Radial outliers: $y_i\sim \mathcal{N}(0,5I_p)$.
\end{itemize}
Point, Cluster, and Radial outliers were introduced by \cite{hubert2012deterministic} and Random outliers by \cite{zhang2023fast}. Distinct from  the settings  of  \cite{hubert2012deterministic} and \cite{zhang2023fast}, here we mix outlier types and use different distances $r$ so that some outliers mask others. We fix $(n,p)=(400,40)$ and apply \Alg{insta} to search over the grid $h=\lfloor 0.5n\rfloor,\lfloor 0.525n\rfloor,\dots,\lfloor 0.975n\rfloor$ with 50 bootstrap pairs.

We consider the following 4 settings:

\begin{itemize}
\item Setting 5: 5\% of cluster outliers with $r=5$.
\item Setting 6: 5\% of point outliers with $r=5$ and 20\% of cluster outliers with $r=50$.
\item Setting 7: 5\% of random outliers with $r=5$ and 15\% of radial outliers with $r=50$.
\item Setting 8: 17.5\% of cluster outliers with $r=5$ and 17.5\% of random outliers with $r=50$.
\end{itemize}
\begin{figure}[tbp]
  \centering
  \includegraphics[width=\textwidth]{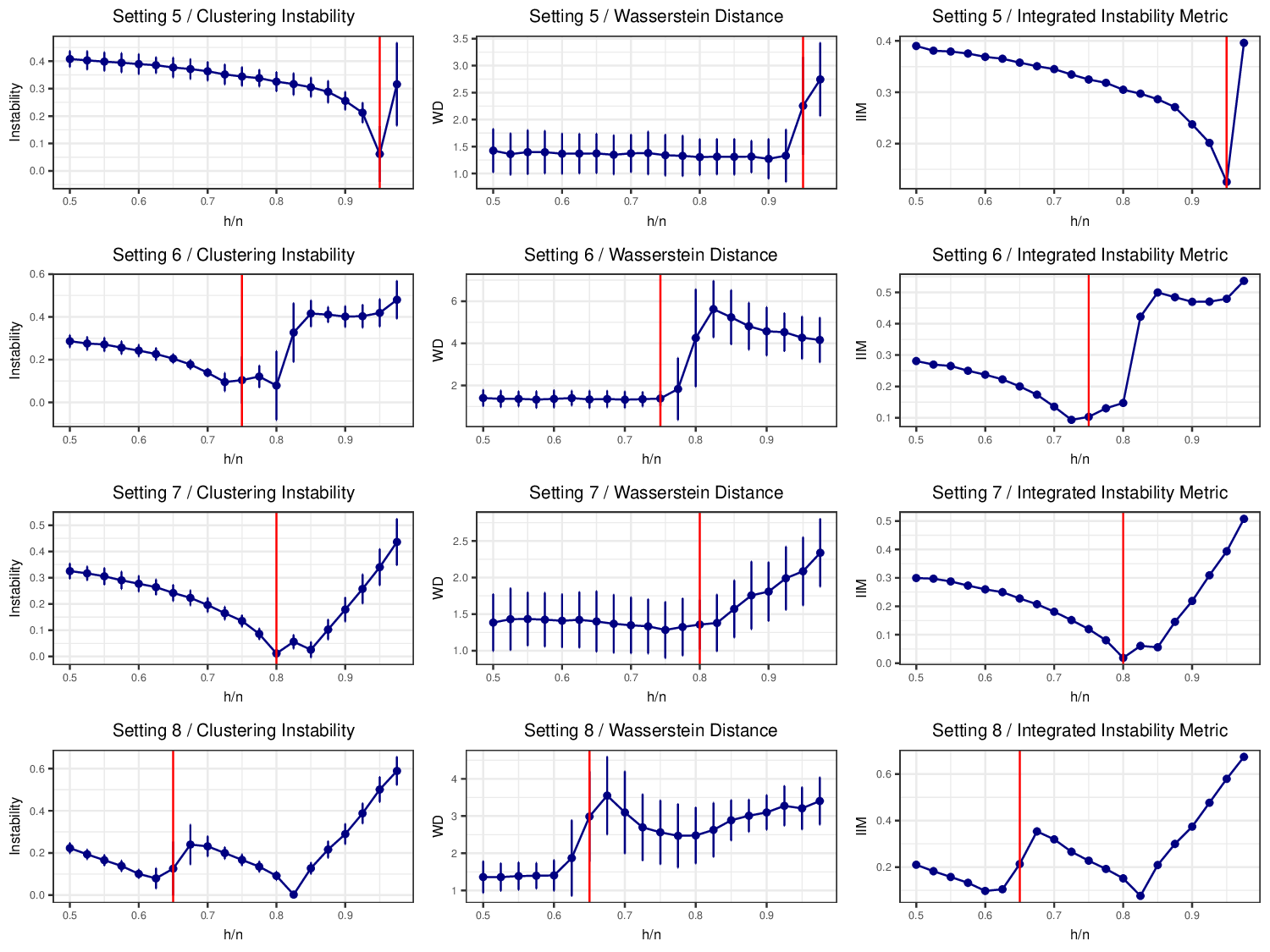}
  \caption{Instability paths for the setting 5,6,7, and 8. The true inlier proportion is highlighted with a red vertical line.}
  \label{fig:hd}
\end{figure}
\Fig{hd} largely mirrors the observations made in \Fig{2Ds}. Masking stymies clustering instability in setting 6 but this failure is overcome by the integrated instability metric. For setting 6, we slightly underestimate the true inlier proportion by 0.025, while in setting 5,7, we correctly recovers the true inlier proportion. 

\begin{figure}[tbp]
  \centering
  \includegraphics[width=\textwidth]{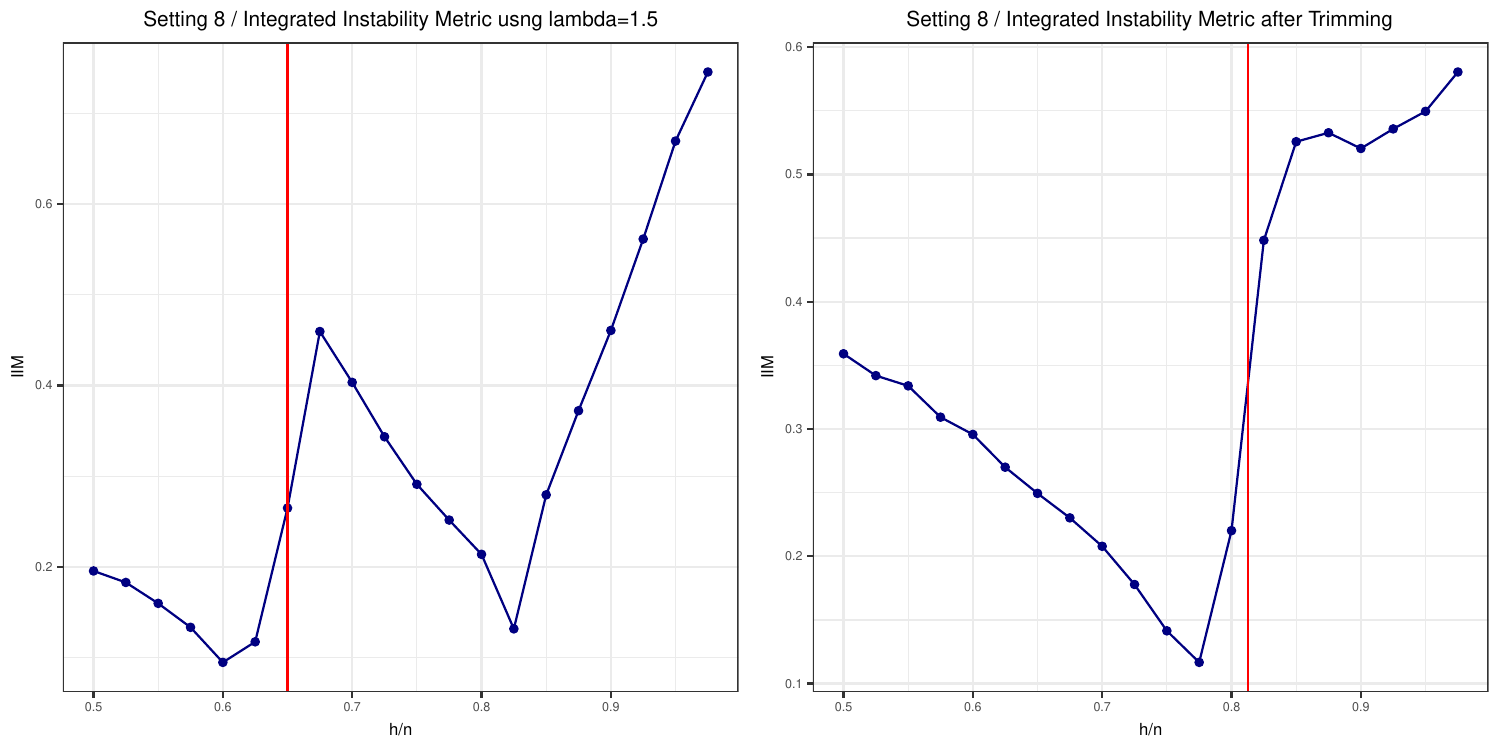}
  \caption{Instability paths for setting 8 using $\lambda=1.5$ and after trimming 20\% of observations. Notice that after trimming, the new ground truth for $h/n$ is 0.8125.}
  \label{fig:masking}
\end{figure}

In setting 8, the clustering instability observed at $\lfloor 0.625n \rfloor$ remains notably high. Although integrating with the log Wasserstein distance reduces the instability gap between $\lfloor 0.625n \rfloor$ and $\lfloor 0.825n \rfloor$, the integrated instability metric still attains its minimum at $\lfloor 0.825n \rfloor$. This implies that the default variance allocation ratio $\lambda=3$ may be insufficient to counteract the masking effect in certain scenarios. Although decreasing $\lambda$ could mitigate this issue, such an adjustment risks exacerbating underestimation bias. As an alternative strategy, trimming extreme outliers using the MCD method and recalculating instability paths proves effective. The left panel of \Fig{masking} demonstrates the integrated instability metric for setting 8 under $\lambda=1.5$, while the right panel displays the IIM after  trimming the top 20\% outliers. We note that in setting 8, the log-Wasserstein distance plot further aids diagnosis: the sharp rise in instability before the 0.7 threshold indicates that the true inlier proportion likely falls below this value. These results highlight the proposed framework’s utility as a diagnostic tool.

The examples in this and the preceding section assume the true inlier count aligns exactly with a subset size in the search grid. While illustrative, this assumption may not hold in practice. For instance, if the true inlier proportion is 0.795, the integrated instability could be minimized at a grid point such as 0.8 due to the discrete nature of the search grid. This discrepancy might result in the instability framework failing to identify certain outliers. To mitigate this issue, a localized grid search can refine the subset size estimation. Empirical observations suggest that, with a sufficiently dense grid, the instability framework tends to favor subset sizes slightly smaller than the true inlier count. 
\section{Real Data Examples}\label{sec:realdata}

\subsection{Star Data}
\begin{figure}[htbp]
  \centering
  \includegraphics[width=\textwidth]{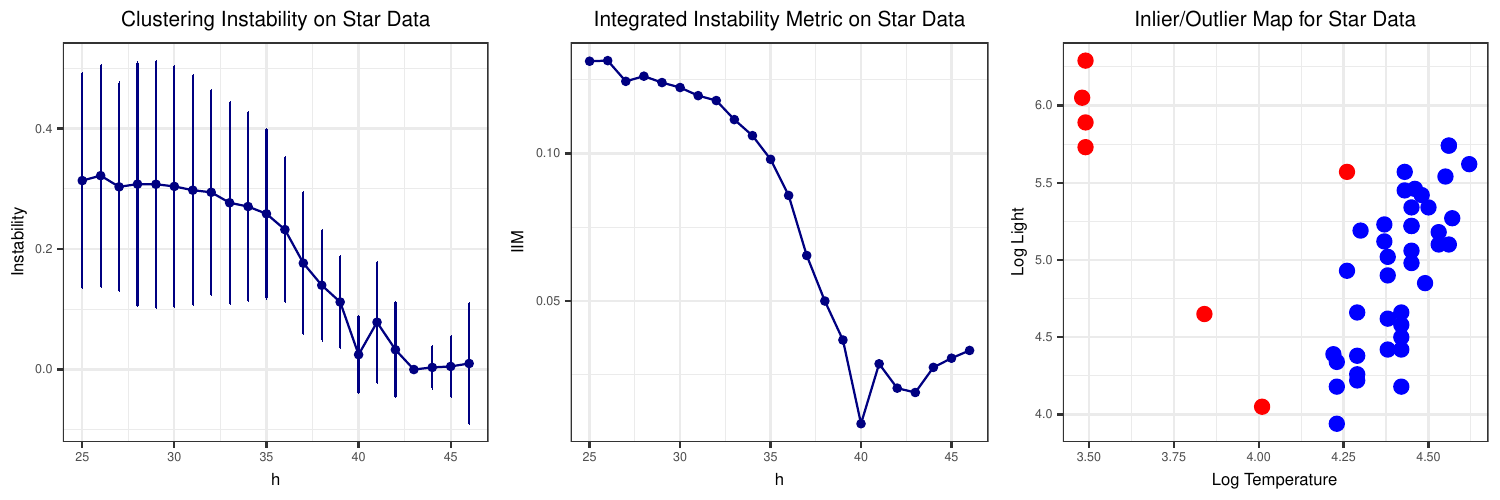}
  \caption{Experimental results for the starsCYG dataset. The left panel shows the clustering instability path. The middile panel displays integrated instability metrics. The right panel depicts inlier/outlier classification using MCD with $h=40$. }
  \label{fig:star}
\end{figure}

Our first real data example is the StarsCYG dataset \citep{rousseeuw2005robust}, which contains 47 logarithmic measurements of light intensity and temperature for the CYG OB1 star cluster. These data were previously analyzed by \cite{berenguer2023model}, who studied the problem of subset size selection in the context of least trimmed squares (LTS) \citep{rousseeuw1984least}. Here we consider the same data but view it through the lens of MCD. Given the small size of the dataset, we can afford to search across all values of $h \in \{25,26,\dots,46\}$ instead of a grid of inlier proportions. The value $h=47$ is excluded because our instability measure is undefined for full data. The grid search takes 3 seconds to complete,  applying \Alg{insta} with 100 bootstrap pairs. From the left panel of \Fig{star}, we observe that clustering instability is minimized at $h=43$.  However, a notable dip in the instability path at $h=40$ suggests the potential presence of a masking effect. The integrated instability metric selects $h=40$, leading us to conclude that there are 7 outliers in total. This conclusion matches the result returned by DetMCD using subset size $\alpha=\lfloor 0.5n\rfloor$, and also $\alpha=\lfloor 0.75n\rfloor$. Notably, this result is more conservative than some existing findings. For instance, \cite{berenguer2023model} identified 5 outliers. 

\subsection{Bank Note Data}
\begin{figure}[htbp]
  \centering
  \includegraphics[width=\textwidth]{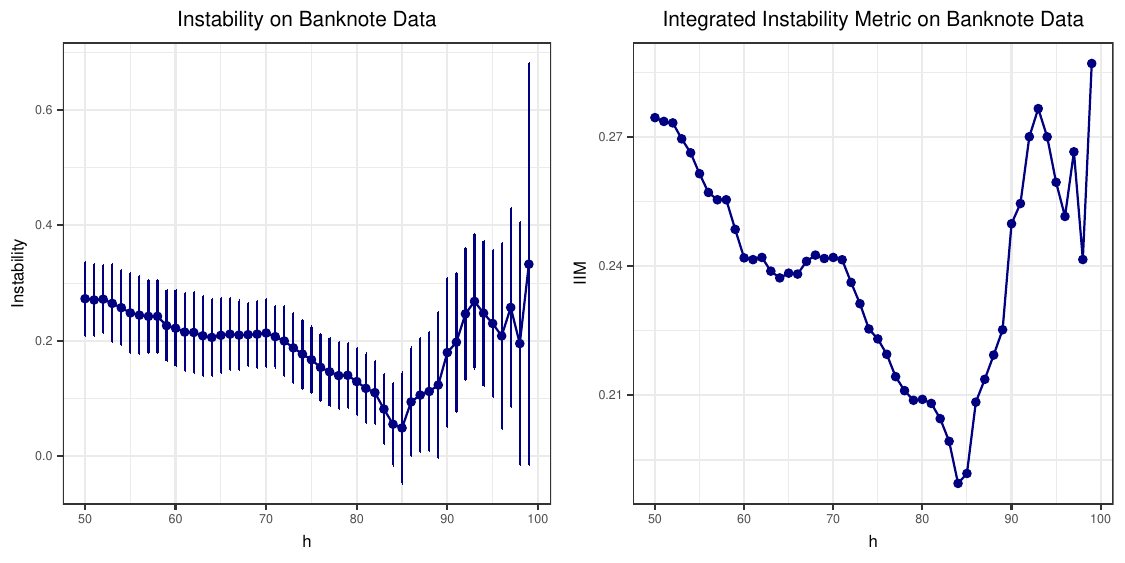}
  \caption{Experimental results for the banknote dataset. The left panel shows clustering instability path, while the right panel shows integrated instability metrics.}
  \label{fig:bank}
\end{figure}

Our second real-world example deals with the forged Swiss bank notes dataset \citep{flury1988multivariate}, which has been previously studied by \cite{hubert2012deterministic} and \cite{zhang2023fast}. This dataset consists of $n=100$ and $p=6$. We now search over $h=\{50,51,52,\dots,99\}$  with \Alg{insta} using 100 bootstrap pairs. The grid search takes 9 seconds to complete. \Fig{bank} demonstrates that the integrated instability metric is minimized at $h=84$, indicating our instability framework identifies 16 outliers. We also apply DetMCD and FDB with $h=75$ and reweighting. DetMCD detects 16 outliers after reweighting, while FDB identifies 19 outliers. The difference arises because DetMCD employs a different cutoff than FDB. We note that the number of outliers identified by DetMCD and FDB depends on the specified subset size. Under the setting $\alpha=\frac{h}{n}=0.5$, DetMCD identifies 21 outliers, and FDB identifies 20 outliers.

\subsection{Fruit Data}\label{sec:fruit}

Our third real-data example, referred to as the fruit data, comprises spectra from three distinct cantaloupe cultivars, labeled D, M, and HA with sample sizes of 490, 106, and 500, respectively. Originally introduced by \cite{hubert2004fast} and later examined by \cite{hubert2012deterministic}, the dataset  includes $1096$ total observations recorded across 256 wavelengths. \cite{hubert2004fast} noted that the cultivar HA encompasses three distinct groups derived from various illumination conditions. Unfortunately, the assignment of individual observations to specific subgroups and the potential impact of the subgroups on spectra are unavailable.
 
\begin{figure}[htbp]
  \centering
  \includegraphics[width=\textwidth]{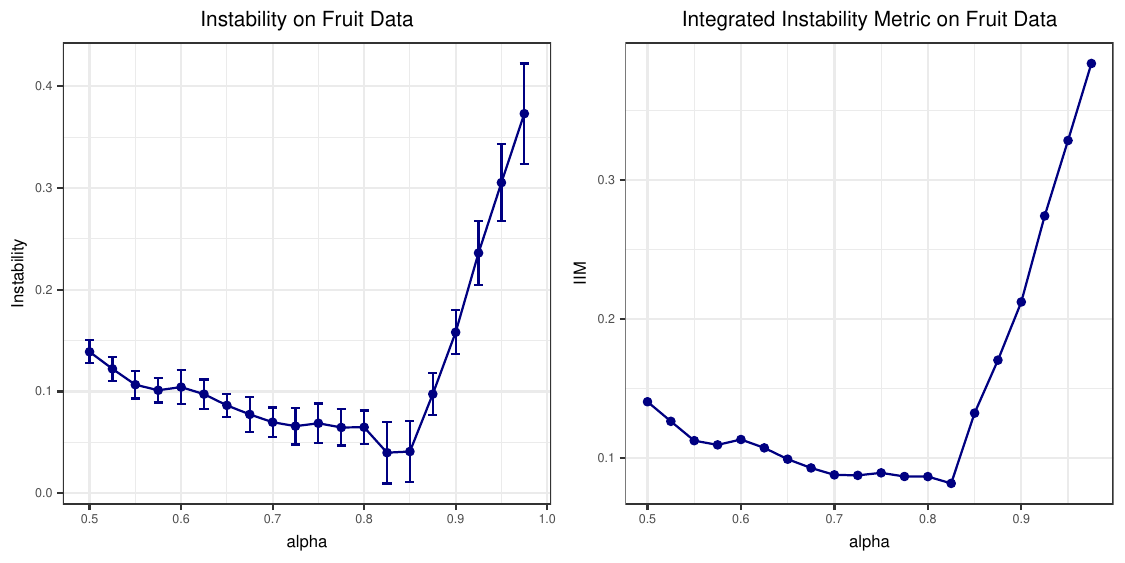}
  \caption{Experimental results for the fruit dataset. Left panel shows clustering instability path, while right panel shows integrated instability metrics. }
  \label{fig:fruit}
\end{figure}

We apply \Alg{insta} and search over the grid $h=\lfloor 0.5n \rfloor,\lfloor 0.525n \rfloor,\dots,\lfloor 0.975n \rfloor$  with 50 bootstrap pairs. The grid search takes 437 seconds to complete. \Fig{fruit} shows the clustering instability path and the integrated instability metrics. The IIM is minimized at $\lfloor 0.825n \rfloor$, so our instability framework suggests that there are around 17.5\%  outliers. We note that the clustering instability is also minized at $\lfloor 0.825n\rfloor$, but the contrast between $\lfloor 0.825n \rfloor$ and $\lfloor 0.85n \rfloor$ is more clear in the right panel. Upon examining the identified outliers, 2 are of the D cultivar, 1 is of the M cultivar, while the remaining 189 are all of the HA cultivar. Combining this with prior information, we speculate that the outliers are mostly from a group measured under different illumination conditions. 

 For this dataset, DetMCD fails to return results due to $n<5p$.  FDB estimates 456 outliers when $h=\lfloor 0.75n\rfloor$ and 496 outliers when $h=\lfloor 0.5n\rfloor$. However, we believe that the results obtained through reweighting are unreliable due to the high collinearity of the data, which results in extreme ill-conditioning of the covariance matrix. In contrast, our instability framework is unfazed by ill-conditioning, as it does not depend on a parametric approximation of Mahalanobis distances. Instead, it relies on the relative ranking of the distances. 

\subsection{Glass Data}\label{sec:glass}
\begin{figure}[htbp]
  \centering
  \includegraphics[width=\textwidth]{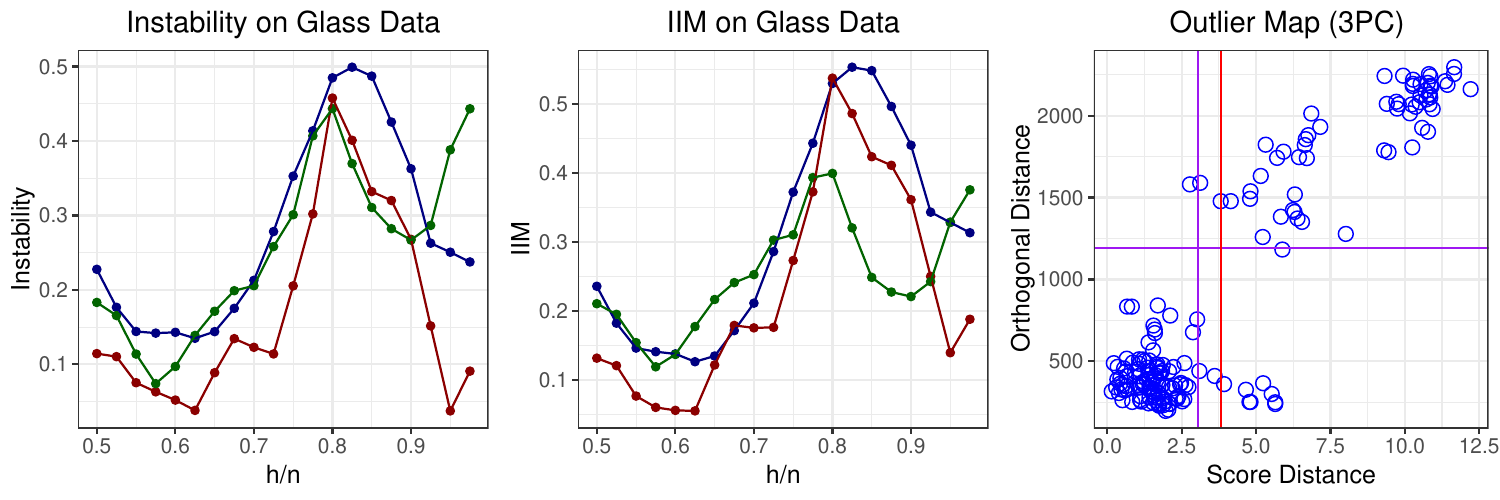}
  \caption{Experimental results for the glass spectra dataset. In the left and middle panels, the blue, red, green lines represent the instability paths at $q=2$, $q=3$, and $q=10$, respectively.}
  \label{fig:glass}
\end{figure}
Our fourth real-data example consists of EPXMA spectra over $p = 750$ wavelengths collected from 180 different glass samples  \citep{Lemberge2000}.  These data were previously analyzed via robust principal component analysis by \cite{hubert2005robpca}.  We apply \Alg{insta2} with 50 bootstrap pairs and search over the two-dimensional grid defined by $h \in \{\lfloor 0.5n \rfloor,\lfloor 0.525n \rfloor,\dots,\lfloor 0.975n \rfloor\}$ and $q\in \{2,3,10\}$. The grid search takes 75 seconds to complete.  Among the three integrated instability metric paths, the minimum is attained at $q=3$ and $h=\lfloor 0.625n\rfloor$. The number of principal components favored by instability matches the choice of \cite{hubert2005robpca}. Their rationale for choosing 3 principal components is that it explains 99\% of the variance (or 96\% of the robust variance). However, \cite{hubert2005robpca} used $h=\lfloor 0.7n\rfloor$ as the subset size, contradicting our instability findings, which suggest 37.5\% high-SD points alone. 

To further investigate the issue, we applied ROBPCA to the glass data using $q=3$ and $\alpha \in \{0.7,0.5\}$. When $\alpha=0.7$, \texttt{PcaHubert} identifies 48 high-SD points and 44 high-OD points, for a total of 51 outliers. Obviously, many of the high-SD points and high-OD points overlap. When $\alpha=0.5$, \texttt{PcaHubert} identifies 73 high-SD points, 62 high-OD points, and 73 outliers in total. Here, all high-OD points are also high-SD points. Thus, we believe that the analysis in \cite{hubert2005robpca} may  have been affected by outlier masking. This highlights the benefit of instability in revealing the number of outliers. The right panel of \Fig{glass} shows the score distances and orthogonal distances returned by \texttt{PcaHubert} for $(q, \alpha)= (3, 0.5)$ and the corresponding cutoff values (purple lines). The figure also indicates the 62.5th percentile of score distances using a red vertical line, which acts as a decision boundary in our nonparametric framework. Given $(q, \alpha)= (3, 0.5)$, the number of high-SD points flagged by \texttt{PcaHubert}  indeed closely aligns with $\lfloor 0.375n\rfloor$.

\subsection{Breast Cancer Data}\label{sec:brca}
\begin{figure}[htbp]
  \centering
  \includegraphics[width=\textwidth]{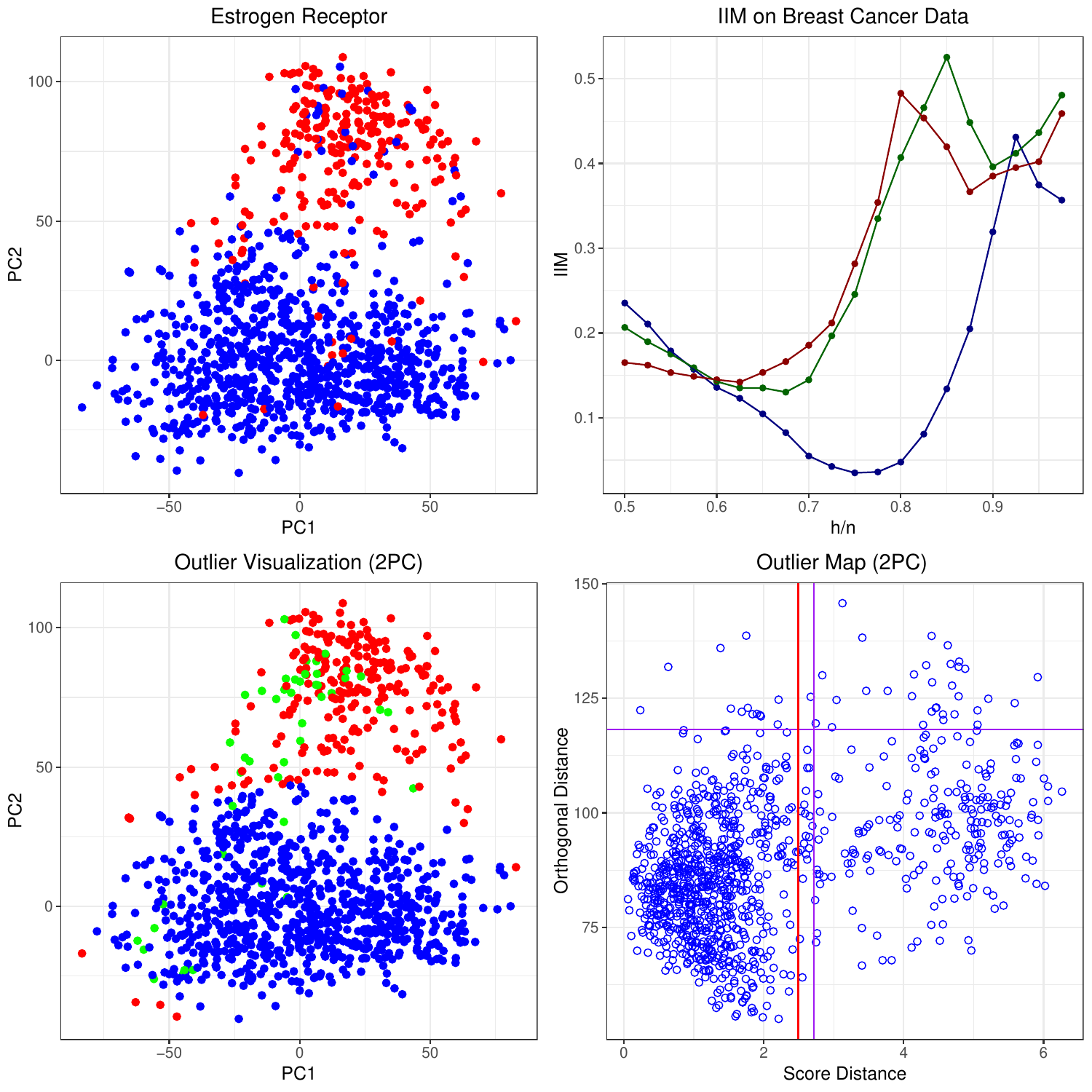}
  \caption{Experimental results for the breast cancer dataset. In the top left panel, blue points are estrogen receptor positive, and red points are negative. The top right panel displays the integrated instability metric paths for $q=2$ (blue), $q=5$ (green), and $q=10$ (red). In the bottom left panel, the blue points are inliers, the red points are high-SD points, while the green points are high-OD points. If a point is both high-SD and high-OD, it is shown in green. In the bottom right panel, the purple lines are cut-offs returned by \texttt{PcaHubert}, while the red line indicates the 75\% percentile of score distance.}
  \label{fig:breast}
\end{figure}
Our final real-data example comes from the The Cancer Genome Atlas (TCGA) project \citep{cancer2012comprehensive}. The breast cancer project (TCGA-BRCA) encompasses  approximately 1,100 patients with invasive carcinoma of the breast. The data are obtained from cBioPortal at \url{https://www.cbioportal.org/} \citep{cerami2012cbio}. We focus on the mRNA expression profiles of the patients. These profiles represent expression levels for 20,531 genes on the 1,100 samples. After a $\log_2(x+1)$ transformation on the expression levels, we retained the top 2,000 most variable genes. Apart from expression profiles, the data also record the estrogen receptor (ER) status of each sample. Women with estrogen receptor negative status (ER-) are typically diagnosed at a younger age and have a higher mortality rate \citep{descriptive}. Of 1,100 samples, 812 are estrogen receptor positive, 238 are negative, 48 are indeterminate, and 2 are missing this information. We retain the 1,050 samples for which the estrogen receptor status is either positive or negative. Thus, our preprocessed data matrix has dimensions $n=1,050$ and $p=2,000$. At this analysis stage, estrogen labels are ignored.

Our search via \Alg{insta2} over the two-dimensional parameter grid $q \in \{2,5,10\}$ and $h\ \in \{\lfloor 0.5n \rfloor,\lfloor 0.525n\rfloor,\dots,\lfloor 0.975n\rfloor\}$ uses
50 bootstrap pairs and takes 1695 seconds to complete. Based on the proportion-of-robust-variance-explained, \texttt{PcaHubert} selects $10$ principal components, the maximal number of principal components allowed. In fact, on these data, the first two principal components  explain only about 23\% of robust variance. \Fig{breast} depicts the results of our analysis. The top right panel of \Fig{breast} indicates that the integrated instability metric favors $q=2$ over $q=5$ and $q=10$. The best parameter combination ($q=2$ and $h =\lfloor 0.75n\rfloor)$ nearly reproduces the proportion of high-SD points found by ROBPCA for $q=2$. The key advantage of the instability framework is that it offers a clear criterion for deciding between different values of $q$.
ROBPCA lacks such a mechanism.  

\begin{table}[htbp]
\centering
\begin{tabular}{@{}ccccccc@{}}
\toprule
            & $q=2$ (I) & $q=2$ (O) & $q=5$ (I) & $q=5$ (O) & $q=10$ (I) & $q=10$ (O) \\ \midrule
ER Positive & 753          & 59      & 746 & 66      & 722           & 90             \\
ER Negative & 21           & 217     & 42 & 196     & 31            & 207            \\ \bottomrule
\end{tabular}
\caption{Confusion matrix for outlier detection at $q=2$, $q=5$, and $q=10$. The symbol I is short for inlier, while the symbol O is short for outlier.}
\label{tab:confuse}
\end{table}

The labels of estrogen status buttress our argument that $q=2$ is a better choice for outlier detection than $q=5$ and $q=10$. To identify outliers when $q=2$, we first select high-SD points nonparametrically by applying MCD with $h=\lfloor 0.75n \rfloor$ to the robust principal component scores. Then we combine these outliers with the high-OD points returned by \texttt{PcaHubert}. At $q=5$ and $q=10$, we directly use the high-SD and high-OD points returned by \texttt{PcaHubert} as outliers. The bottom left panel of \Fig{breast} depicts our outlier classification at $q=2$. Here, we find 273 outliers in total, while at $q=5$ and $q=10$, we find 262 and 297 outliers in total. \Tab{confuse} shows the confusion matrix with ER negative as the ``positive" and minority class. Outlier detection accuracy is maximized at $q=2$. Compared with $q=2$, both the numer of false positives and false negatives increase at $q=5$ and $q=10$, with $q=10$ having the worst overall performance.

\section{Discussion}\label{sec:discussion}

This paper introduces a novel method for model selection within the minimum covariance determinant (MCD) framework. We propose bootstrap clustering instability as a tool to uncover the distribution of inliers and outliers and to determine an appropriate number of inliers, $h$. To address the challenges posed by masking outliers, we combine clustering instability with the log Wasserstein distance to define an integrated instability metric. Additionally, combining our stability framework with robust principal component analysis allows us to identify the number of principal components that most effectively distinguish between low- and high-score distance points.

Our approach is rigorously evaluated against well-established variants of the MCD algorithm, including the deterministic minimum covariance determinant (DetMCD), the fast depth-based (FDB) algorithm, and robust principal component analysis (ROBPCA). In simulations, our instability framework excels in accurately estimating the number of inliers and improves the estimation of location and scatter parameters by retaining a larger fraction of the data. Real-world datasets further demonstrate the effectiveness of instability paths for outlier detection, yielding consistent and insightful results.

In practice, we recommend visualizing all three different types of instability paths. Combining numerical results with visual evidence maximizes the potential of the instability framework and minimizes the risk of error. We believe the proposed framework is a valuable addition to the existing robust statistics toolbox, with the potential for broader applications across other robust estimation tasks.

\begin{center}
    {\large \textbf{Supplementary Materials}}
\end{center}
\begin{description}
\item[Supplement:] A pdf file that contains proof of \Thm{distance}, a new computational technique for FastMCD, and additional simulation results. 
\item[Software:]  R package of the described method, along with scripts to recreate \Fig{2Ds}, \Fig{hd}, and \Fig{masking}.  An online repository is also available at \url{https://github.com/qhengncsu/StableMCD}, where the real-data examples are fully reproducible.
\end{description}
\section*{Acknowledgements}
We are grateful to the two anonymous referees for their insightful comments, which significantly improved both the presentation and methodology of this article. We also thank the associate editor and editor for their prompt handling of the manuscript. In particular, we are indebted to one referee for identifying the critical masking issue, which directly motivated the development of the related solutions presented in this work. We further thank Seyoon Ko and Do Hyun Kim for their valuable suggestions. This research was supported in part by USPHS grants GM53275 and HG006139.

\bibliographystyle{asa}
\bibliography{refs}

\end{document}